\documentclass[
preprint,
showpacs,
amsmath,amssymb,
aps,
pre,
]{revtex4-1}

\usepackage{amsmath}
\usepackage{amsthm}

\usepackage{enumerate}
\usepackage{subfig}

\usepackage{empheq}

\usepackage[dvipsnames]{xcolor}

\usepackage{graphicx}% Include figure files
\usepackage{dcolumn}% Align table columns on decimal point
\usepackage{bm}% bold math
\usepackage{hyperref}% add hypertext capabilities
\usepackage[utf8]{inputenc}
\usepackage[english]{babel}
\usepackage[notref,notcite,final]{showkeys}
\RequirePackage[normalem]{ulem}
\RequirePackage{color}

\begin{document}

\title{Multiple bubble dynamics and velocity selection in Laplacian growth without surface tension}
		\author{Mark Mineev-Weinstein}
		\email{mark\_mw@hotmail.com}%
		\affiliation{New Mexico Consortium, Los Alamos, NM, 87544, USA; \email{mark\_mw@hotmail.com}}
\author{Giovani L.~Vasconcelos}
		\email{giovani.vasconcelos@ufpr.br}%
		\affiliation{Departamento de F\'{\i}sica, Universidade Federal do Paran\'a, 81531-990 Curitiba, Paran\'a, Brazil}	

\begin{abstract}
A new selection phenomenon in nonlinear interface dynamics is predicted. A generic class of exact regular unsteady multi-bubble solutions in a Hele-Shaw cell is presented.  These solutions show that the case where the asymptotic bubble velocity, $U$, is twice greater than the velocity, $V$, of the uniform background flow, i.e., $U = 2V$, is the only attractor of the dynamics. Contrary to common belief, the predicted velocity selection requires neither surface tension nor other external regularization.
\end{abstract}

% \begin{keyword}
% {Hele-Shaw flow} \sep { Laplacian growth} \sep {Free boundary problem} \sep  {Conformal mapping} \sep {Multiply connected domain}
% %\MSC[2010] 00-01\sep  99-00
% \end{keyword}

%
%\date{\today}

%\linenumbers
\pacs{47.20.Hw, 47.20.Ma, 47.15.km, 02.30.Ik}

\maketitle

\section{Introduction}

Nonlinear interface dynamics and pattern formation out of equilibrium were and still are at the forefront of fundamental physics and remain also  of great importance for numerous applications \cite{Pelce}. The Laplacian growth (LG), that is, interface motion under potential flow, is probably one of the deepest and at the same time the simplest and the most universal process of this kind, as it grasps all major features of unstable interface dynamics:

--  nonlinearities;

-- instabilities;
 
-- formation of universal patterns;

-- multiple applications, ranging from oil/gas recovery to malignant growth.

\noindent 
It also stands as the prototype for many growth processes, such as  dendritic solidification \cite{Langer}, combustion fronts \cite{Pelce}, electro-migration of voids \cite{void}, streamer ionization fronts \cite{streamer}, and bacterial colony growth \cite{bacteria}, to name just a few.  \noindent In addition, it has remarkable hidden integrable structure behind its mathematical formulation (see below). 

The recent surge of  interest in LG was motivated by the set of remarkable discoveries of  deep connections of the LG problem with  branches of fundamental physics, such as quantum gravity \cite{MWZ} and quantum Hall effect \cite{ABWZ}.  In addition, strong connections of LG with {\it mathematics} were revealed,  on both  the {\it classical} (the inverse potential problem  \cite{GTV}, Riemann surfaces \cite{KMWZ})  and {\it modern} levels (integrable hierarchies \cite{KKMWZ}, random matrices \cite{MPT}, and the soliton theory of the hydrodynamic kind \cite{KMWZ}).  

The major challenge of LG interface dynamics in a Hele-Shaw channel (HSC), posed by Saffman and Taylor in 1958 \cite{ST}, 
was the selection of a single observable pattern (mainly a finger or a bubble) from a continuum of stationary solutions obtained in the absence of surface tension  (ST).  Since then, this problem has been at the center of attention 
because of highly non-trivial physics and mathematics that defied traditional treatments. 
Since the problem was presented in \cite{ST}, the invariably held common belief was that it is ST that selects the observable pattern.  This belief was  confirmed when, after considerable efforts in the 1980s, it was finally shown by several groups  \cite{SurfaceTension1,SurfaceTension2,SurfaceTension3,SurfaceTension4} that the inclusion of small ST indeed leads to the observable velocity selection. After these works  the selection problem was considered to be resolved.

Quite surprisingly, it was shown later, first in \cite{98PRL} for a finger and later in \cite{us2014, withArthur, Robb2015} for a bubble,  that there is no need of ST (nor any other external regularization) for velocity selection in a HSC. In  these works the velocity is selected entirely within the zero surface-tension (ZST) dynamics,  by demonstrating that the selected pattern is the only attractor of the dynamical system describing the  {interface} evolution. Since these works addressed only a single interface,  it was not clear if selection without ST would hold for multiple interfaces. 
This article aims precisely to address selection for multiple interfaces. (Laplacian growth in multiply connected domains was considered earlier  in \cite{Richardson} and  \cite{KMWZ}, albeit in a different context.)

In \cite{us2014} we conjectured (a) that our ZST selection mechanism works for an arbitrary number of bubbles,  and  (b) that  all bubbles reach  the same asymptotic velocity, which is precisely twice  the velocity of the uniform background flow. In other words, $U=2V$ in the long time limit,  where $U$ and $V$ are the bubble and background flow velocities, respectively.  Mathematical difficulties to describe  multiple interfaces  prevented us  at that time to verify  these two conjectures. 

In this paper, we extend  the  exact  ZST  steady solutions for  multiple interfaces in a HSC reported in \cite{GLV2015} to  the {\it unsteady} case.   By using these new unsteady solutions, we demonstrate
that any number of bubbles  reach the same asymptotic velocity, $U=2V$, thus proving a conjecture stated in \cite{us2014}.
This result contributes to both physics and mathematics.  In physics, it predicts a novel effect, easily experimentally testable \cite{swinney} and important for 
numerous applications.
In mathematics, it presents a new rich class of unsteady exact solutions  for a notoriously challenging problem---unstable nonlinear dynamics with infinite degrees of freedom. Thus, it also contributes to the field  of integrable nonlinear systems --- an intensively developing subdiscipline of mathematical physics. 

A similar class of regular ZST solutions  for a finger  \cite{SLPRL98}  faithfully reproduces the real dynamics observed in numerics  in the low ST limit. But bubble dynamics in a Hele-Shaw cell is much less understood than finger evolution (see \cite{okumura} for open problems in bubble dynamics).
It is both tempting and important  for science and  applications to obtain exact solutions against which to compare experiments and simulations to remove any controversies (mostly induced by instabilities).

Below, we derive a 
rich class of  unsteady non-singular multi-bubble ZST solutions.  Obtaining these solutions is possible due to the integrability of the LG problem, which allows one to  recast the problem as a set of {\it conservation laws}. The solutions are given in terms of a conformal mapping from a canonical multiply connected domain (interior of the unit circle with a set of smaller disks removed) to the fluid region outside the bubbles.  The  mapping is written in closed form in terms of certain special transcendental functions --- the secondary Schottky-Klein prime functions \cite{PRSA2014}. 
The Schottky-Klein prime functions were first used  \cite{Crowdy2005} to construct conformal mappings for multiply connected domains, and  applied to several problems \cite{CrowdyReview}, including steady multiple bubbles in an unbounded Hele-Shaw cell  \cite{Crowdy2009}.
% and an unsteady fluid blob in a HSC  \cite{CrowdyTanveer}. 
Later, the   secondary Schottky-Klein prime functions were introduced \cite{PRSA2014}, which allowed to address multiple steady bubbles in a HSC \cite{GLV2015}. Here we shall use the  secondary Schottky-Klein prime functions to construct exact unsteady solutions for multiple bubbles in a HSC and  study  velocity selection  in this context.

To address the velocity selection problem in the asymptotic regime, we need only solutions  that stay regular for all times. Thus, we exclude initial data  leading to  solutions that cease to exist   or lose univalence in  finite time.  In the theory of ill-posed problems,  such  data  narrowing is known as Tikhonov regularization \cite{tikhanov, tikhanov_arsenin, ivanov,lavrentev, kabanikhin}. {Then,  the problem becomes well-posed, provided 
the data left after such  removal  are 
continuous w.r.t.~all smooth initial data.}
This restricted set of data is called the {\it set of  well-posedness} of the problem.   {Within the set of well-posedness,  one deals only with regular solutions.}
After the  Tikhonov regularization, one can study all aspects of this dynamical system,  including the problem of selection {\it without surface tension or any other external regularization}.

The paper is organized as follows.  We formulate the problem in Sec.~\ref{sec:formulation}  and obtain explicit unsteady solutions  in Sec.~\ref{sec:exact}.  In Sec.~\ref{sec:selection}, we analyze the solutions for $t \to \infty$  and establish  $U=2V$ as the only attractor of the ZST dynamical system; we also review other selection mechanisms in the same section. All these results are summarized in Sec.~\ref{sec:conclusion}.

\section{Mathematical formulation} 

\label{sec:formulation}

\subsection{Multiple bubbles in a Hele-Shaw channel}

\begin{figure}
\centering 
\subfloat[\label{fig:1a}]{\includegraphics[width=0.6\textwidth]{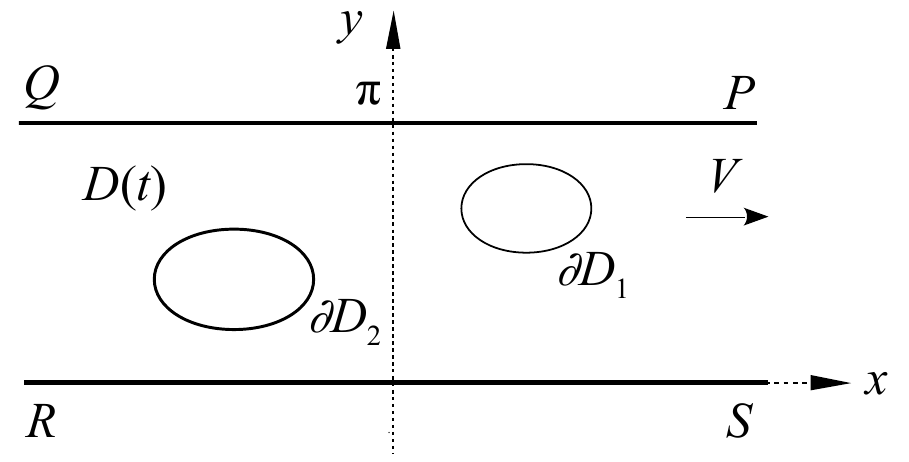}} \\
\subfloat[\label{fig:1b}]{\includegraphics[width=0.6\textwidth]{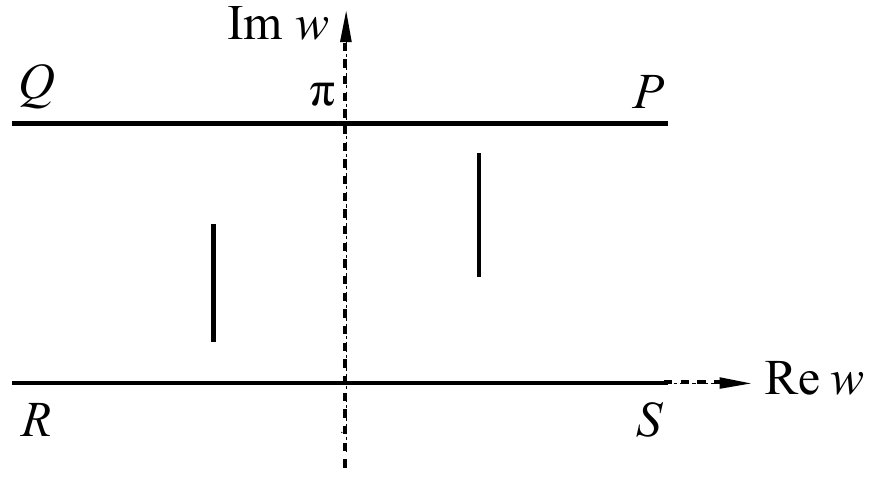}} \\
\subfloat[\label{fig:1c}]{\includegraphics[width=0.6\textwidth]{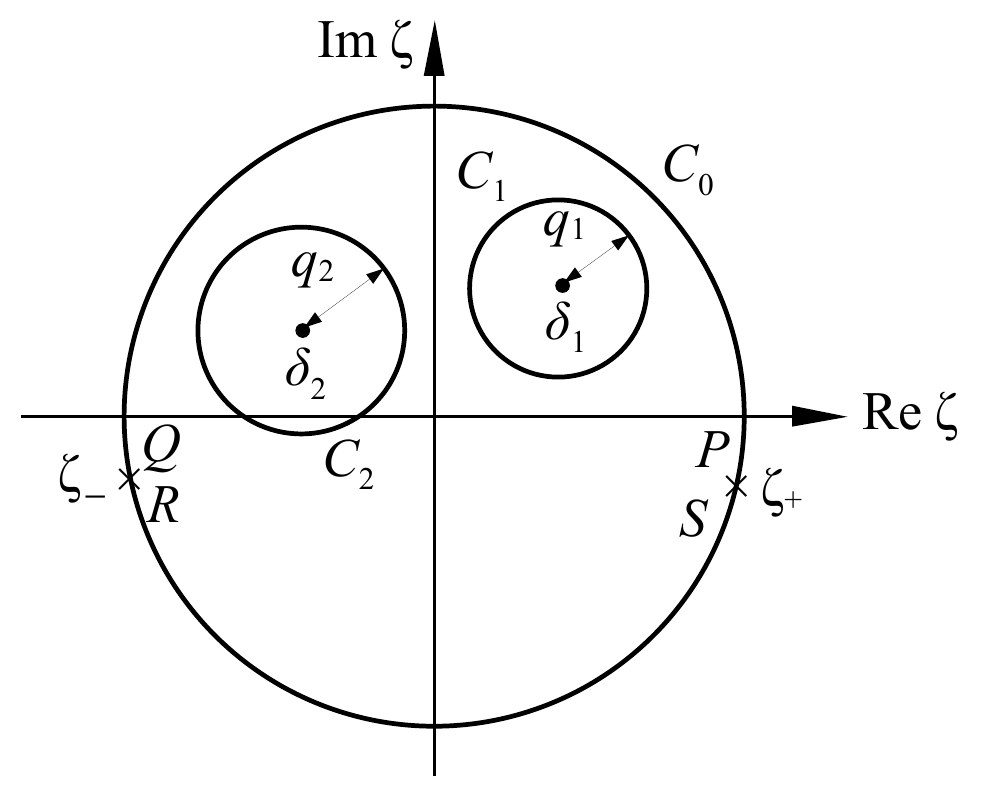}} 
\caption{The flow domains for a finite assembly of bubbles in  a Hele-Shaw channel: (a)  the physical $z$-plane; (b)  the $w$-plane of the complex potential;  and (c)  the auxiliary complex $\zeta$-plane. The points $\zeta_\pm$ in the $\zeta$-plane are mapped  to $x=\pm\infty$ in the $z$-plane.}
\label{fig:1}
\end{figure}

The Hele-Shaw channel is an infinite (in both directions) horizontal strip occupied by a viscous fluid (oil), denoted by the domain $D(t)$, and by $M$ inviscid bubbles, represented by the non-intersecting domains $D_j(t)$, $j=1,2, ..., M$,  trapped inside the oil, as shown in Fig.~\ref{fig:1a} for $M=2$. The uniform oil source at $x = -\infty$ pushes the flow toward the uniform sink at $x = +\infty$.  The problem is formulated (in scaled units where the channel width is chosen to be $\pi$ and the background velocity  $V$ is set to unity) as follows. As oil is incompressible,  i.e.,~${\rm div}~{\bf v} = 0$, where ${\bf v}$ is the oil velocity vector field, and  the flow is potential, ${\bf v} = - ~{\rm grad}~p$ (Darcy's law), where $p$ is pressure, we obtain the 2D Laplace equation: $\nabla^2 p = 0$. It also follows from above that $\lim_{x \to \pm \infty} p = -x$.
Also, $\partial_n p = 0$ (no normal flow) at the channel boundaries, $y = 0$ and $y=\pi$, where $\partial_n$ is the normal derivative. In addition, we have $p=p_j(t)$ along  the moving interface  $\partial D_j(t)$, for $j=1,...,M$, where the $p_j$'s  are constant along the bubble boundaries, since pressure is constant inside the  bubbles and   ST is neglected. Finally, the kinematic boundary condition requires that the normal velocities of the moving boundary, $V_n$, and of the oil, $v_n$, coincide, so $V_n = -\partial_n p$ at the moving interface $\Gamma(t)=\bigcup_j \partial D_j$. The full mathematical formulation of the problem thus takes the following form:
\begin{subequations}
\label{eq:1}
\begin{empheq}[left=\empheqlbrace]{align}
	& \nabla^2 p = 0 \qquad \quad \mbox{in}\, D(t) ,  \label{eq:1a}\\
	 &  p=p_j    \qquad\qquad \mbox{at}\  \partial D_j(t) ,  \label{eq:1b} \\
	 & V_n = -\partial_n p  \quad \quad \mbox{at}\  \Gamma(t), \label{eq:1c} \\
	 & p = -x \qquad \quad  \  \mbox{for}\  x \to \pm \infty,  \label{eq:1d}\\
	 & \partial_n p = 0 \qquad \quad \  \mbox{at}\ y = 0\ {\rm and}\  y=\pi.\label{eq:1e}
\end{empheq}
\end{subequations}

The goal is to find closed form solutions of (\ref{eq:1}) for the unsteady dynamics of an {\it arbitrary}  number of inviscid droplets. Earlier solutions  \cite{us2014}   were restricted to a single interface. 
To deal with multiple bubbles we must recast the problem in terms of conformal mappings between multiply connected domains.
Let us note that Laplacian growth in multiply connected domains is not  completely specified by the initial configuration only  \cite{KMWZ}, since it is necessary to prescribe additional conditions from which one can determine the pressures $p_j$. Rather than specify {\it a priori} the pressures $p_j$, we shall instead fix certain geometric parameters of the conformal mapping, so as to ensure that the  $p_j$'s  remain constant in time;
see Sec.~\ref{sec:Une2}.

%\cite{footnote0}. 

\subsection{The conformal mapping}

\label{sec:map}

Let us solve (1a)--(1e) 
using a conformal mapping $z(\zeta,t)$ from a special
domain in an auxiliary complex $\zeta$-plane to the oil domain $D(t)$.
Since $\nabla^2 p=0$  we  introduce a complex potential $w(z,t)=-p(x,y;t)+i \psi(x,y;t)$, analytic in $D(t)$  and subject  to  boundary conditions indicated below. 

All three complex planes, $z$, $\zeta$, and $w$, mentioned above, are  shown in Fig.~\ref{fig:1} for the case $M=2$ (i.e., two bubbles).  The physical plane $z=x+ iy$, see Fig.~\ref{fig:1a},  represents the oil with $M$ bubbles, $D_j(t)$ ($j=1,...,M$), moving to the right together with the oil.
The oil domain, $D(t)$, occupies the horizontal strip, $0<y<\pi$, excluding the bubbles regions, $ D_j(t)$.  The complex  potential $w$-plane is shown in Fig.~\ref{fig:1b}.  
The vertical slits there represent the bubble boundaries, $\partial D_j(t)$, and the horizontal lines describe the north/south walls of the channel:
\begin{empheq}[left=\empheqlbrace]{align*}
	&{\rm Re} ~ w =  -p =  -p_j,\cr
	 &  {\rm Im} ~ w  = \psi = 0, \pi ,
\end{empheq}
where the first condition indicates isobars in accordance with (\ref{eq:1b}) and the second one stands for streamlines in accordance with (\ref{eq:1e}).

The complex plane, $\zeta = \xi + i\eta$, used for  the conformal map $z(\zeta,t)$,  is shown in Fig.~\ref{fig:1c}. The oil domain $D_\zeta$  in the $\zeta$-plane corresponds to the interior of the unit circle $C_0$, with  $M$ disjoint circles, $C_j$, $j=1,...,M$, excised from it.  Here $\delta_j$ and $q_j$ are the center and the radius of the $j$-th circle. The circles $C_j$ are mapped to the bubble boundaries, $\partial D_j(t)$, in the $z$-plane and to the vertical slits in the $w$-plane, whereas  $C_0$ is mapped to the channel walls, $y =0, \pi$, in the $z$-plane and to $\psi = 0,\pi$ in the $w$-plane, while the  points $\zeta_\pm\in C_0$   are respectively mapped to the sink and source at $x = \pm \infty$.
For uniqueness of conformal mapping between two given domains, we fix the following three real parameters: 
 $\delta_1=0$, so that the circle $C_1$ is concentric with $C_0$,  and  $\zeta_+=1$. (We shall, however,  keep the symbol $\zeta_+$  in the formulas below for convenience of notation.)

Let us double  the flow domain $D(t)\to D(t)\cup \overline{D(t)}$, where bar denotes complex conjugation, 
so adding a new set of bubbles  within this extended  domain, which are the mirror reflections of 
$D_j(t)$ with respect to the real axis. 
Similarly, the extended domain in the $\zeta$-plane, to be denoted by 
$F_0$, is obtained by adding to $D_\zeta$ its reflection in the unit circle $C_0$:
\begin{equation}
F_0=D_\zeta\cup\varphi_0(D_\zeta),
\label{eq:F0}
\end{equation}
where $\varphi_0(\zeta) = 1/\overline\zeta$ denotes reflection in $C_0$.
The planar domains $D_\zeta$ and $\varphi_0(D_\zeta)$
can be seen as the two halves of the Riemann surface (Schottky double) \cite{KMWZ,Schiffer}  obtained by ``gluing'' each circle $C_j$ to its reflection, $C_{-j}=\varphi_0(C_j)$, in $C_0$ (see \cite{Crowdy2005, PRSA2014}).

For later use, let us  introduce the following Moebius maps:
\begin{equation}
\theta_j(\zeta)={\delta_j}+\frac{q_j^2 \zeta}{1-\overline{\delta}_j\zeta}, \qquad j=1,...,M,
\label{eq:theta}
\end{equation}
whose inverses are 
\begin{equation}
\theta_{-j}(\zeta)\equiv \theta_j^{-1}(\zeta)= \frac{1}{\overline{\theta}_j(1/\zeta)},
\label{eq:thetainv}
\end{equation}
where  $\overline f(\zeta)$ denotes the Schwarz conjugate of $f(\zeta)$ defined by  $\overline f(\zeta)=\overline{f\left(\overline \zeta\right)}$.
One can verify that $\theta_j(\zeta)$ maps the interior of the circle $C_{-j}$ onto the exterior of  the circle  $C_j$. The set $\Theta_0$ consisting of all compositions of the maps $\theta_j$ and their inverses defines  a classical Schottky group  and the region $F_0$ defined in (\ref{eq:F0}) is a fundamental region of the group $\Theta_0$ \cite{Baker}.

Of particular interest to us here is the subgroup $\Theta_M\subset\Theta_0$ that consists of only and all {\it even} combinations of  the maps $\theta_j$, $j=1,...,M$, and their inverses.  
For some $1\le l \le M$, let us then consider the set of  $2M-1$ M\"obius maps,  $\psi_k$, defined by
\begin{align}
\psi_k\equiv
\theta_l  \circ\theta_{k}, \qquad  k=-M,...,M, \ k\ne 0,-l.
\label{eq:psi}
\end{align}
These maps  are  fundamental generators of the subgroup $\Theta_M$,  and  the same group $\Theta_M$ is generated irrespective of the choice of  $l$ \cite{PRSA2014}. Associated with the Schottky group $\Theta_M$, one can  define certain special transcendental functions, known as the  secondary Schottky-Klein prime functions $\Omega_M(\zeta,\alpha)$, see Appendix A, in terms of which the solution for $z(\zeta, t)$ will be given in Sec.~\ref{sec:exact}.

\subsection{Symmetry relations} 

\label{sec:S}

Equation (\ref{eq:1e}) implies that  the conformal map $z(\zeta,t)$  satisfies the following boundary condition:
\begin{align}
{\rm Im}\,[z(\zeta,t)]= \mbox{constant}, \quad \mbox{for}\quad \zeta\in C_0.
\label{eq:Imz}
\end{align}
Using that $\overline \zeta=1/\zeta$ for $\zeta \in C_0$,  we can recast   (\ref{eq:Imz}) as
\begin{align}
 \overline{z}(1/\zeta,t) = z(\zeta,t)+\mbox{constant} \qquad \mbox{or} \qquad   \overline{z}(1/\zeta,t)\simeq z(\zeta,t),
\label{eq:barz}
\end{align}
where   $h(z) \simeq f(z)$ means that $h(z) = f(z) + \mbox{constant}$.
Equation (\ref{eq:barz}) is valid for $\zeta\in C_0$ and elsewhere by analytic continuation. Another symmetry relation for $z(\zeta,t)$ can be obtained by considering the boundary conditions at the bubble interfaces, as shown below.

We note that equations (\ref{eq:1a})--(\ref{eq:1c}) are equivalent to   
\begin{equation}
{\cal S}_t = 2w_z,
\label{eq:St}
\end{equation}
where subscripts are partial derivatives and ${\cal S}(z,t)$ is the Schwarz function \cite{Davis,  Howison92} of the curve $\Gamma(t)$, defined as $\overline z = {\cal S}(z,t)$ for $z \in \Gamma(t)$.  It follows from (\ref{eq:St})  that all singularities of ${\cal S}(z,t)$ in $D(t)$ different from those of $w(z)$ are constants of motion \cite{Howison92}.

One can show that the Schwarz function of the $j$-th bubble boundary $\partial D_j$  has the following representation in the $\zeta$-plane
\begin{align}
	 g_j(\zeta,t)\equiv {\cal S}_j(z(\zeta,t),t) = z(\theta_{-j}(\zeta),t )
		\label{eq:S}
\end{align}
(see  Appendix B). 
The  Schwarz functions $g_j(\zeta,t)$, $j=1,...,M$, differ only by constants, 
\begin{align}
g_k(\zeta, t)=g_j(\zeta, t) + G_{kj},
\label{eq:glm}
\end{align}
where $G_{kj}$  may depend on time but not on $\zeta$. In other words, the  Schwarz functions ${\cal S}_j(z,t)$ computed from each individual boundary $\partial D_j$ can be seen as different logarithmic branches of the  global Schwarz function ${\cal S}(z,t)$. Now, from (\ref{eq:St}) and the boundary conditions on $w(z,t)$, it follows that  ${\cal S}_j(z,t)$  has constant imaginary part on  the channel walls or alternatively: ${\rm Im}[g_j(\zeta,t)]=\mbox{constant}$ for $\zeta\in C_0$.  This, together with  (\ref{eq:barz}) and (\ref{eq:S}), implies  (see Appendix B)  that 
\begin{align}
	 %g_j(\zeta,t) = z(\theta_{j}(\zeta),t ) +\mbox{constant} \qquad \mbox{or} \qquad  
	 g_j(\zeta,t) \simeq z(\theta_{j}(\zeta),t ).
		\label{eq:Sj}
\end{align}

From (\ref{eq:S})--(\ref{eq:Sj}) we then obtain the following important symmetry relations:
\begin{align}
	 z(\psi_k (\zeta),t) \simeq z(\zeta,t ) ,
\label{eq:zpsi}
\end{align}
where $\psi_k = \theta_l\circ \theta_k$, for a fixed $l$ and $k=-M, ..., M$, with $k\ne 0$. Since the maps $\psi_k(\zeta)$ are  generators of the group $\Theta_M$, see (\ref{eq:psi}), Eq.~(\ref{eq:zpsi}) states that $z(\zeta,t)$ is invariant (up to an additive constant) under each element of $\Theta_M$. In  other words, $z(\zeta,t)$ is an {\it additive automorphic function}  \cite{Ford} with respect to the group $\Theta_M$. This property is  crucial to constructing exact solutions to the problem of multiple bubbles in a HSC, as will be seen next.

\section{Exact time-dependent solutions}

\label{sec:exact}

In what follows, we address {\it non-singular} solutions, which remain finite for all times. Among other advantages, these solutions enable us to solve the selection problem in the long-time asymptotics. In other words, we leave aside solutions that blow up in finite time, either by developing cusps or losing univalence, as physically non-realizable.  In this  way we regularize this initially ill-posed problem, that is, we  convert it into a well-posed one (see more details in Sec.~\ref{sec:Une2}). But first, we would like to present a rich class of  {\it unsteady} non-singular multi-bubble solutions  by extending  earlier results   \cite{98PRL} (fingers) and  \cite{us2014} (single bubble). This class describes all smooth initial data with any desirable degree of accuracy. 
%It is well known by virtue of the classical Weierstrass theorem that polynomials constitute a complete basis to approximate any smooth function with any required accuracy \cite{weierstrass}. 
(It is  well known that ratios of polynomials, i.e., rational functions or so-called Pad\'e  approximants, provide a complete description of all smooth curves %\old{and is even much more efficient than approximation by polynomials}
\cite{baker}. 
%\old{Because} 
Since linear combination of logarithms, which is our solutions, is 
%\old{just}
an integral of rational functions with simple poles, whose sum give the most generic Pad\'e approximants, this explains why our solution given by (\ref{eq:z0}) describes any smooth curve as close as desired.)

\subsection{The complex potential}

The following formula links the complex potential $W(\zeta,t)\equiv w(z(\zeta,t),t)$ and our {\it key} function, $\Omega_M(\zeta, \alpha)$ (see (\ref{eq:SKM}) and \cite{GLV2015}):
\begin{equation}
W(\zeta,t) = \log \frac{\Omega_M(\zeta, \zeta_-) \, \Omega_M(\zeta, \theta_l(\zeta_+))}{\Omega_M(\zeta,\zeta_+) \, \Omega_M(\zeta, \theta_l(\zeta_-))} + ic_l(t),
\label{eq:W}
\end{equation}
where $l$ can be $1, 2, ..., M$, and $c_l(t)$ is a real constant \cite{footnote_cl}. Using the properties of $\Omega_M(\zeta, \alpha)$  given in  Appendix A (or in \cite{GLV2015}), one can verify that $W(\zeta,t)$ satisfies  the desired boundary conditions (\ref{eq:1b}) and (\ref{eq:1e}), which in this case read: 
\begin{empheq}[left=\empheqlbrace]{align*}
	&{\rm Re}[W(\zeta,t)]=-p_j=\mbox{constant}, &\mbox{for $\zeta\in C_j$},
	\cr
	 &  {\rm Im}[W(\zeta,t)] = \mbox{constant},  &\mbox{for $\zeta \in C_0$}.
\end{empheq}
If $\zeta$ in $C_l$ we find that
\begin{align}
p_l=\log \left|\frac{ \delta_l-\zeta_+}{ \delta_l -\zeta_-}\right|.
\label{eq:pj}
\end{align} 
We shall henceforth set $l=1$,  which implies $p_1=0$, since $\delta_1=0$.  Thus, the pressure on the  bubble surface $\partial D_1$ is taken as the reference pressure with respect to which the pressures in the other bubbles are measured. Furthermore, setting $\delta_1=0$ in (\ref{eq:theta}) yields $\theta_1(\zeta)=q_1^2\zeta$.

\subsection{Exact Unsteady  Solutions} 
\label{sec:Une2}

As we want to study selection in the asymptotic regime, we must deal exclusively with solutions $z(\zeta, t)$  that remain non-singular for all times.  This restriction to initial data leading to non-singular solutions is known as Tikhonov regularization of ill-posed problems, whereby the well-posedness of the problem may be  {ensured}  by narrowing the class of possible initial data to an appropriate set, called the {\it set of  well-posedness} of the problem \cite{ivanov,lavrentev}. By  extending non-singular solutions from simpler geometries [17-19] to the case of multiple bubbles,  we have found that the solutions in all these cases 
have the same form: $z(\zeta, t)$ is the sum of time-dependent {\it logarithms} (see below). 

In terms of the Schotky-Klein formalism,  the solution  for $z(\zeta,t)$ takes the form
 \begin{align}
z(\zeta, t) = & h(t)  +i \Delta +\log \frac{\Omega(\zeta, \zeta_-)}{\Omega(\zeta, \zeta_+)} 
+ \alpha_0 \log \frac{ \Omega(\zeta,   q_1^2 \zeta_+)} {\Omega(\zeta,   q_1^2 \zeta_-)} \cr
&+\sum_{k=1}^N \left[ \alpha_k \log  \Omega(\zeta,   a_k) + \overline \alpha_k \log \Omega(\zeta, 1/\overline a_k)\right],
\label{eq:z0}
\end{align}
where  the subscript from $\Omega_M$ has been dropped to save space.
Here the function $h(t)$ and constants  $\Delta$ and $\alpha_0$ are real (with  $|\alpha_0|<1$), $a_k(t) \notin F_0$ are complex  [see (\ref{eq:F0}) to recall the definition of the fundamental region $F_0$], and  the complex constants $\alpha_k$ satisfy 
\begin{align}
\sum_{k=1}^N \alpha_k = 0,
\label{sumalpha}
\end{align}
 to ensure that $z(\zeta,t)$ is single-valued. Without loss of generality, the values of $a_k(0)$ are chosen to be inside the circle $C_1$ and outside the circle obtained by the inversion of $C_0$ in $C_1$, i.e., $q_1^2<|a_k|<q_1$.  The quantity $\Delta$  in  (\ref{eq:z0})  can be  determined from the condition that   ${\rm Im}[z(\zeta,t)]=0$  for $\zeta$ on the lower segment of the unit circle, as shown in  Fig.~\ref{fig:1} \cite{footnote2}.
 
When $M=1$, Eq.~(\ref{eq:z0})  reproduces a single bubble dynamics   \cite{us2014}. In this case, the circular domain $D_\zeta$ becomes an annulus, and the Schottky-Klein prime functions are reduced to Jacobi theta functions \cite{CrowdyReview}, in terms of which the solution for a single bubble was expressed in \cite{us2014}. For more than one bubble ($M>1$) one must employ the full-fledged secondary prime functions.

Making use of properties of  $\Omega(\zeta, \alpha)$ [see Appendix A], one can  verify that $z(\zeta,t)$ given in (\ref{eq:z0}) satisfies the required symmetry relations (\ref{eq:barz}) and (\ref{eq:zpsi}). One  also sees from (\ref{eq:W}) and (\ref{eq:z0}) that $w(z)\approx z$ for $|z|\to \infty$, as required by (\ref{eq:1d}).
Furthermore, the moving boundary of the $j$-th bubble is given by $\partial D_j(t) = z(\delta_j + q_i e^{i\varphi}, t)$, where $\varphi \in [0, 2\pi]$ parametrizes the bubble boundaries.  To find the  time dependency of $a_k(t)$ and $h(t)$ in (\ref{eq:z0}), 
%of of the parameters of the mapping function $z(\zeta,t)$, 
we shall consider a set of conserved quantities of the dynamics.

First let us consider the conserved singularities of the Schwarz function ${\cal S}(z,t)$. From (\ref{eq:Sj}) we obtain that 
\begin{align}
g_1(\zeta, t) =~& h(t)+ i \Delta'+ \log \frac{\Omega(q_1^2 \zeta, \zeta_-)}{\Omega(q_1^2 \zeta, \zeta_+)} + \alpha_0   \log \frac{\Omega(\zeta, \zeta_+)}{\Omega(\zeta, \zeta_-)} \cr
& + \sum_{k=1}^N \left[ \alpha_k \log \Omega(q_1^2\zeta, a_k) + \overline \alpha_k  \log \Omega(q_1^2\zeta,\overline a_k)\right],
\label{eq:sg}
\end{align}
where in writing the term containing $\alpha_0$ we used  that ${\Omega}(q_1^2 \zeta, q_1^2 \alpha)={q_1^2} {\Omega}(\zeta,\alpha)$; see relation  (\ref{eq:id5}) in Appendix A. Here $\Delta'$ is a real-valued quantity whose expression is not relevant for our discussion.
Recalling that the $a_k$'s are inside $C_1$ but outside the reflection of $C_0$ in  $C_1$,  meaning that $q_1^2<|a_k|<q_1$, it  follows from (\ref{eq:sg}) that the only singularities of $g_1(\zeta,t)$ in  $D_\zeta$  are the points $\zeta_k=\varphi_0(a_k)=q_1^2/\overline{a_k}$, for $k=1,...,N$. 
This implies, in turn, that the only singularities of ${\cal S}(z,t)$ in the fluid domain $D(t)$ are  located at the points  $\beta_k$, where 
\begin{align}
\beta_k=z(q_1^2/\overline{a_k},t), \qquad k=1,...,N.
\end{align}
More explicitly, we have
\begin{align}
\beta_k =~ & h(t) + i \Delta + \log \frac{\Omega(q_1^2/\bar a_k, \zeta_-)}{\Omega(q_1^2/\bar a_k, \zeta_+)} 
+ \alpha_0 \log \frac{ \Omega(1/\bar a_k,    \zeta_+)} {\Omega(1/\bar a_k,    \zeta_-)} \cr
&+ \sum_{m=1}^N \left[ \alpha_m \log \Omega(q_1^2/\bar a_k, a_m) + \bar \alpha_m  \log \Omega(q_1^2/\bar a_k, 1/\bar a_m)\right].
\label{betak}
\end{align}

Since all singularities of   ${\cal S}(z,t)$ within  $D(t)$  are constants of motion, we have $\dot \beta_k=0$, where dot denotes time derivative. 
{\it Analytically}, the $\beta_k$'s are the logarithmic, time-independent singularities of the Schwarz function, ${\cal S}(z,t)$, located in $D(t)$, while {\it geometrically} they are vertices of virtual fjords \cite{fjord},  akin to those in \cite{MS94a,MS94b}.

We can obtain a second set of conserved quantities by analyzing the asymptotic behavior of ${\cal S}(z,t)$ for $x\to\pm\infty$. First, it follows from (\ref{eq:sg})   that  if $\alpha_0\ne 0$ then  $g_1(\zeta,t)$ has logarithmic singularities at $\zeta_\pm$:
\begin{align}
g_1(\zeta,t) \approx \pm \alpha_0 \log(\zeta-\zeta_\pm)+ h(t)+B_\pm(t), \qquad \zeta\to\zeta_\pm,
\end{align} 
so   ${\cal S}(z, t)$ behaves for $x\to\pm \infty$ as
 \begin{align}
{\cal S}(z,t)= -\alpha_0 z +  \frac{2}{U} h(t)+ C_\pm(t)+O\left(\frac{1}{z}\right), 
\label{eq:S2}
\end{align}
where 
\begin{align}
U=\frac{2}{1+\alpha_0}.
\label{eq:U}
\end{align}
 and $C_\pm(t)$ remain finite for all times. (The time-dependency of $B_\pm(t)$ and $C_\pm(t)$ are irrelevant for the present discussion.) 
From $\lim_{|x|\to\infty} w_z =  1$ and (\ref{eq:St}), it follows that $ \lim_{|x|\to\infty} {\cal S}_t(z,t) = 2$, so by virtue of (\ref{eq:S2})  the quantities
\begin{align}
\beta_\pm =  \frac{2}{U} h(t) -2t +  C_\pm(t)
\label{betapm}
\end{align}
are constant in time, i.e., $\dot \beta_\pm=0$.
One can show that ${\rm Im}[\beta_+]= {\rm Im}[\beta_-]$, hence there are only three independent real quantities for the two complex constants $\beta_\pm$.

The total area $A$ of the bubbles can be neatly expressed through the constants $\beta_k$ and $\beta_\pm$ as
\begin{align}
\frac{A}{\pi}= \frac{\beta_+ -\beta_-}{2} +  {\rm Re} \sum_{k=1}^N \overline {\alpha}_k \beta_k  ,
\label{eq:A}
\end{align}
and so $A$ is  guaranteed to remain constant throughout the evolution.  However,  we do not have analytical formulas for the areas, $A_j$, of the individual bubbles.

As discussed in \cite{KMWZ},  multi-bubble LG  requires $M-1$ additional relations to determine the pressures $p_j$, for $j=2,...,M$ (recall that we have set $p_1=0$ as a reference pressure). One can   fix either  the bubble pressures $p_j$ or the bubble areas $A_j$. In either case we have  $2N+M+2$  conserved real quantities:    $N$ complex  $\beta_k$ ($k=1,...,N$), $\beta_\pm$ (recalling that ${\rm Im}[\beta_+]= {\rm Im}[\beta_-]$),  and $A_j$ (or $p_j$), for $j=2,..,M$. We can enforce these conditions by keeping $\delta_j$, for $j=2,..,M$, fixed in time and solving for the remaining $2N+M+2$ real parameters: $N$ complex parameters  $a_k$ and  $M+2$ real parameters, namely $h$,  $\gamma_-=\arg(\zeta_-)$, and $q_j$, $j=1,...,M$. (Recall that we have used the three degrees of freedom of the Riemann mapping theorem to set $\delta_1=0$ and  $\zeta_+=1$.)
As explicit formulas for $A_j$ and $p_j$ are either not available  or hard to deal with numerically, we  fix both $\delta_j$ and $q_j$, for  $j=2,...,M$,  in time.  Then we are left with $2N+3$ real free parameters: $N$ complex  $a_k$ and real $\gamma_-$, $h$, and $q_1$. There is equal number of  conserved real quantities:   $N+2$ complex  $\beta_k$ and  $\beta_\pm$, subject to ${\rm Im}[\beta_+]= {\rm Im}[\beta_-]$.

The conservation laws (\ref{betak}) and (\ref{betapm}) provide (implicitly) the full dynamics  of the conformal mapping $z(\zeta,t)$, thus completing the description of  generic unsteady motion for  multiple  bubbles.  Initial conditions should be chosen carefully to avoid {possible} finite-time singularities due to  formation of cusps or loss of univalence (in short, to avoid finite time blow ups). Fortunately, there is a large set of such initial data. A simple rule to choose such favorable initial data is to place the $\beta_k$'s either behind the bubbles or clearly out of the bubbles' reach, so that these points are all left behind in the long-time limit and the solution approaches a steady regime. (If a bubble gets too close to one of the $\beta_k$'s the interface might eventually self intersect, which means that the solution blows up in finite time.)
We note, however, that determining the bubbles' evolution for a given initial data requires solving numerically a set of nonlinear algebraic equations with transcendental special Schottky functions --- a  difficult task that we plan to address in the near future.
In this context, it is perhaps worth mentioning that in the simpler case of a single bubble, where the solutions are written in terms of well-known elliptic functions, several numerical examples of interface evolution have been reported \cite{withArthur}.

After we have presented the full dynamics in Eqs.~(\ref{betak}) and (\ref{betapm}), let us now focus on the long-time limit ($t\to+\infty)$ in order to address the velocity selection --- the main goal of this article.  

\subsection{Long-time Asymptotics}
\label{sec:asym}

Let us now consider the asymptotic behavior of the solution (\ref{eq:z0})  for $t\to\infty$. From (\ref{betapm}) one sees that in order to keep $\beta_\pm$ fixed in time we must have $h(t)\to Ut$, for $t \to \infty$, since all other terms remain finite in this limit. Similarly, the divergence of $h(t)$  must be cancelled in (\ref{betak}) by a divergent logarithmic term, since $\beta_k$ must remain constant for all times. This divergence cancelation  can only arise from the third term in the RHS of (\ref{betak}),  so we need $q_1^2/\bar a_k \to \zeta_-$ for $t\to\infty$.  
This  implies that  all $a_k$'s must approach the point  $q_1^2\zeta_-$ when $t \to +\infty$, and  they do so exponentially slow:
\begin{align}
a_k(t)=q_1^2\zeta_-+c_ke^{-\lambda_k t},
\end{align}
where ${\rm Re}[\lambda_k]>0$ and $c_k$ are asymptotic constants. (The expressions for $\lambda_k$ and $c_k$ are too lengthy and irrelevant for the present argument.)

In this limit and since $\sum_k \alpha_k=0$,  the solution (\ref{eq:z0}) reduces to
\begin{equation}
z(\zeta, t) = Ut + \log \frac{\Omega(\zeta, \zeta_-)}{\Omega(\zeta, \zeta_+)} + \left(1-\frac{2}{U}\right) \log \frac{\Omega(\zeta, q_1^2\zeta_-)}{\Omega(\zeta, q_1^2\zeta_+)},
\label{eq:Une2}
\end{equation}
where  a nonsignificant  additive constant was omitted. In other words, in the long-time asymptotics the assembly of $M$ bubbles reaches a steady regime where all bubbles move to the right with the velocity $U$ \cite{GLV2015}.  This result holds regardless of the  conditions imposed to determine the pressures $p_j(t)$, as it is based only on the conserved quantities $\beta_k$ and $\beta_\pm$, defined in (\ref{betak}) and (\ref{betapm}).

%We have thus seen that if we start with an initial multi-bubble configuration whose Schwarz function ${\cal S}(z,0)$ has a simple pole at infinity, i.e., $\alpha_0\ne 0$ in (\ref{eq:z0}), then in the asymptotic limit $t\to\infty$ the system reaches a steady  regime where the bubbles   move  with velocity $U\ne 2$, with the final velocity $U$ being determined  by the strength of the pole at infinity; see (\ref{eq:S2}).  If, however, the initial configuration is such that ${\cal S}(z,0)$ is analytic at infinity, i.e., $\alpha_0=0$, then the bubbles will reach a steady solution with $U=2$. We shall see  furthermore  in Sec.~\ref{sec:asymp} that  the steady solutions with $U\ne 2$ are unstable with respect to any shape perturbation that displaces the initial singularity of ${\cal S}(z,0)$ from infinity.

We have thus succeeded to extend the steady solutions for multiple bubbles obtained in \cite{GLV2015} to the {\it unsteady} regime, which is a crucial step to address velocity selection.

\section{Velocity selection} 

\label{sec:selection}

\subsection{Selection for fingers and single bubbles: Brief review}

As mentioned in the Introduction, the most challenging aspect of the Hele-Shaw interface dynamics is the {\it selection problem}, posed in 1958 by Saffman and Taylor in their seminal work \cite{ST} on viscous fingering in a HSC. They observed that the asymptotic shape (the finger) reaches a width that is exactly one half of the channel width  (or equivalently the finger velocity $U$ is twice the velocity $V$ of the background flow of the viscous fluid, i.e., $U=2V$). In the same work \cite{ST},  a continuous family of stationary solutions for all values $0 < U < \infty$ was found. Then why only the value $U=2V$ from this continuous family is observed (or selected)? The answer to this problem, conjectured in \cite{ST}, attributed the observed selection to surface tension (ST), which is neglected in the continuum family of exact solutions. But the authors of \cite{ST} were unable to confirm this conjecture because of
mathematical difficulties to include ST. 

The reason of these difficulties (which also occurred in subsequent attempts of this kind) is that ST, here denoted by $\sigma$, is a singular perturbation in this problem. In other words, the value $\sigma =0$ is an essential singularity of the mapping function $z(\zeta,t)$. Therefore this function cannot be expanded as a power series of ST near zero, and so ST cannot be used for a 
regular perturbation treatment. 

In 1984 Kruskal and Segur \cite{Kruskal-Segur-1} found a remedy to overcome this difficulty by treating (exponentially) small ST corrections of the finger shape. They developed  ``asymptotics beyond all orders''  \cite{Kruskal-Segur-2}, which  is a variation of the well known quasiclassical (also called WKB) approach in quantum mechanics. The role of a small Planck constant $\hbar$ in WKB is played by (also small) $\sigma$ in the selection problem. Using this  approach several groups \cite{SurfaceTension1,SurfaceTension2,SurfaceTension3,SurfaceTension4} showed in the mid 1980s that inclusion of ST switches the continuous spectrum mentioned above to a discrete countable family, all of which converge to $U=2V$ when ST approaches zero. The same approach was also applied to a single {\it bubble} in a HSC \cite{tanveer86,tanveer87} whereby it was shown that the same velocity $U=2V$ is  selected. With these works, the long-standing conjecture that ST causes selection was finally confirmed and the beyond-all-orders asymptotic analysis was accepted as the solution to the selection problem. The dynamical selection for multifingers   has been considered in \cite{Paune}. See also \cite{Casademunt} for a  survey (up to 2004) of the selection problem in Hele-Shaw flows.

The commonly accepted wisdom that  ST is necessary  for selection was unexpectedly challenged in 1998, when the observed finger was selected without ST \cite{98PRL}. Using a general class of exact unsteady non-singular solutions, it was possible to show that the steady solution with $U=2V$ is the only attractor of the dynamics. This analysis  \cite{98PRL} demonstrated that the special nature of the solution with $U=2V$ is already present (or ``built in'') in the zero surface tension Laplacian growth, in agreement with selection by  inclusion of surface tension and other regularizations. 
Indeed, in later works  it was theoretically found \cite{king}, and numerically confirmed  \cite{mccue}, that other boundary conditions, such as kinetic undercooling \cite{king}, also lead to the same $U=2V$ selection. More strikingly,  similar selection has  also been observed experimentally in non-fluid systems, such as finger-like ionization fronts in electric breakdown \cite{streamer}, where there is no analog of ST, thus confirming that external regularizations are not necessary for velocity selection. Alternative attempts to get selection without surface tension have  been made by maximizing some function of the pattern shape \cite{Aldushin} and by considering a finite channel  flow \cite{Feigenbaum}, although 
%the later work was criticized in \cite{CrowdyTanveer} where 
it was subsequently argued in \cite{CrowdyTanveer} that finite size effects are irrelevant  for both selection and the interface dynamics far from  boundaries.

More recently,  the velocity selection $U=2V$  was demonstrated (again without ST) for a single bubble  \cite{us2014,withArthur,Robb2015}, using an approach   similar to that used  for a finger  \cite{98PRL}.
In \cite{us2014},  a new family of {\it unsteady} solutions for a single bubble in a channel helped to demonstrate  that  $U=2V$ is the only stable fixed point (attractor) of this  dynamical system.  The same result was obtained for a single bubble in an unbounded cell \cite{Robb2015}. In this work, 
by extending the unsteady dynamics from a single  to an arbitrary number of bubbles, we arrive to  the same selection, $U=2V$, as discussed next.

\subsection{Velocity selection for multiple bubbles}

\label{sec:asymp}

From Eqs.~(\ref{betak}) and (\ref{betapm}) for the full time dynamics of  the singularities it follows that there are two fixed points for the $a_k(t)$'s:   $q_1^2\zeta_-$ (shown to be an attractor in Sec.~\ref{sec:asym}) and $q_1^2\zeta_+$.
%If we view the motion   of the    of the map $z(\zeta,t)$ given in (\ref{eq:z0}) as a  dynamical system, then the discussion in Sec.~\ref{sec:exact} shows ; see, e.g., the steady solution given in (\ref{eq:Une2}).  
%Furthermore, the argument presented in Sec.~\ref{sec:asym} reveals that $q_1^2\zeta_-$ is an attractor of the dynamics. Let us now show that 
The second fixed point $q_1^2\zeta_+$ is, on the contrary,  a repeller, as shown next.

Recall from Sec.~\ref{sec:Une2} that if  we initially  have a map $z(\zeta,0)$ with singularities at $a_0^{\pm}(0)=q_1^2\zeta_\pm$, i.e., $\alpha_0\ne 0$ in (\ref{eq:z0}), 
then such a configuration (provided it exists for all times) will evolve to a steady regime with $U\ne 2$. Let us  now perturb the initial  configuration  by displacing slightly  this  singularities from $q_1^2\zeta_\pm$:  i.e., $a_0^{\pm}(0)=q_1^2\zeta_\pm+\varepsilon_\pm$,  when $|\varepsilon_\pm|\ll 1$, while $a_k(0)$, for $k=1,...,N$, stay intact. Then  the term containing $\alpha_0$ in (\ref{eq:z0}) takes the  form 
%\begin{widetext}
\begin{align}
\frac{\alpha_0}{2}  \log \frac{ \Omega(\zeta,   q_1^2 \zeta_++\varepsilon_+)\Omega(\zeta,   1/(q_1^2 \zeta_+^ {-1}+\overline{\varepsilon_+}))} {\Omega(\zeta,   q_1^2 \zeta_-+\varepsilon_-)\Omega(\zeta,   1/(q_1^2 \zeta_-^{-1}+\overline{\varepsilon_-}))},
\label{eq:a0}
\end{align}
which is chosen  to preserve the symmetry relations  (\ref{eq:barz}) and (\ref{eq:zpsi}). Under the above perturbation, initial singularities at the fixed points $q_1^2\zeta_\pm$ are shifted to non-fixed points $q_1^2\zeta_\pm+\varepsilon_\pm$, which must then move to the point $q_1^2\zeta_-$ for $t \to +\infty$ (see Sec.~\ref{sec:asym}), implying that
%\begin{subequations}
%\label{eq:1}
\begin{empheq}[left=\empheqlbrace]{align*}
	&\varepsilon_+\to  q_1^2 (\zeta_--\zeta_+),\cr
	 &  \varepsilon_-\to 0 ,
\end{empheq}
or more completely
%\end{subequations} 
\begin{align}
a^\pm_0(t)=q_1^2\zeta_\pm+\varepsilon_\pm(t)= q_1^2\zeta_-+O(\exp(-\lambda_\pm t)), \qquad t\to +\infty,
\end{align}
where ${\rm Re}\,\lambda_\pm >0$. Thus, the term (\ref{eq:a0}) approaches zero asymptotically. 

Since all other singularities $a_k(t)$ also move to  $q_1^2\zeta_-$,   the  sum in (\ref{eq:z0})  vanishes,  so we arrive at the following asymptotic solution:
\begin{equation}
z(\zeta, t) = 2t + \log \frac{\Omega(\zeta, \zeta_-)}{\Omega(\zeta, \zeta_+)}.
\label{eq:U2}
\end{equation}
Thus,  the bubble assembly reaches the velocity $U=2$ in the steady regime.
This result establishes that the only stable fixed point (attractor) for the $a_k(t)$'s describing an  arbitrary number of moving bubbles in the channel  corresponds to $U=2$.

\section{Conclusion}

\label{sec:conclusion}

We have presented a new class of exact solutions  (\ref{eq:z0}) for unsteady motion of an assembly of inviscid bubbles in a Hele-Shaw channel when  surface tension is neglected. The solution was constructed in terms of a conformal mapping from a canonical multiply connected  domain  to the viscous fluid region outside the bubbles.  This mapping was obtained in closed form in terms of the secondary Schottky-Klein prime function introduced in \cite{PRSA2014}.
In this way, we solved in quadrature the initial value problem for multiple  bubbles in a channel.

This means that we have converted the differential equations for the problem   (\ref{eq:1a})--(\ref{eq:1e})
to algebraic-like equations for the constants of motion (\ref{betak}) and (\ref{betapm}) which fully govern the dynamics of the mapping singularities $a_k$'s. Then, given initial values of all $a_k$'s ($k=0,1,...,N$) in the set of well-posedness of the problem (see above), our solutions  (\ref{eq:z0}) remain non-singular for all times and thus describe the multi-bubble evolution for $0\le t<\infty$. 
Since any initial shape can be expressed in terms of our logarithmic class of solutions with any desirable accuracy (because  logarithms are integrals of rational functions with simple poles which can approximate any smooth function), it means that we solved the problem exactly and   completely. 
This became possible because  of integrability of our non-linear initial value  problem. 
Finding specific numerical solutions for a given initial data is a more challenging problem that is left  for the future, as it is outside the scope of the present  work. We note in passing, however, that  in the simpler case of a single bubble several numerical examples have been reported \cite{withArthur} where the solutions display many of the features discussed here in the more general multibubble setting.

After obtaining the full multi-bubble dynamics, we analyzed the long-time regime ($t\to +\infty$), in order to address the velocity selection problem --- the main physical result of our paper.
(The details of the pre-asymptotic dynamics is an interesting topic for future work.)
We have  shown that  in the long-time asymptotics the assembly of  bubbles  reaches the same velocity, $U = 2V$, which is precisely twice the background uniform flow velocity, $V$, provided by the source (and sink) at $|x| \to \infty$.  Our results thus show that within the generalized class of non-singular solutions,  the steady solution  with $U=2V$ is the only attractor of the time-dependent dynamics.
This confirms our conjecture stated earlier in \cite{us2014} that $U=2V$ should be selected irrespective of the number of bubbles.  The predicted phenomena is expected to be observed  in experiments and probably cannot be described analytically by more traditional mathematical tools.

It is instructive to compare the dynamical selection mechanism discussed above with the standard approach to the selection problem via regularization of the Saffman-Taylor problem by  including surface tension. In the latter,  it was shown via asymptotics beyond all orders that only a discrete set of steady solutions survive the regularization procedure and that all such solutions converge to $U=2V$ when  surface tension approaches zero. This selection  by surface tension was limited to the steady-state framework, thus leaving aside the  pre-asymptotic stages; while our ZST selection describes the whole evolution of arbitrary interfaces for  $0\le t<\infty$. Also, by revealing the  instability  (without surface tension) of all continuum of steady solutions except for $U=2V$, we significantly clarified the 
dynamical origin of velocity selection in LG. Here we have extended  for multiple interfaces the approach originated in \cite{98PRL,us2014} for a single interface.

G.L.V.~is grateful to the Department of Mathematics at the Imperial College in London, where this work was initiated. M.M-W.~acknowledges the Brazilian agency CAPES for financial support and thanks both the Department of Physics at UFPE, Brazil, and the Simons Center for Geometry and Physics for hospitality during intermediate stages of the work. G.L.V.~acknowledges partial funding from Conselho Nacional de Desenvolvimento Cient\'ifico e Tecnol\'ogico (Brazil) under Grant No.~312985/2020-7.

\section*{Appendix A: The Schottky-Klein prime functions}

Associated with the Schottky subgroup $\Theta_M$ introduced in Sec.~\ref{sec:map}, one can  define the  secondary Schottky-Klein prime functions $\Omega_M(\zeta,\alpha)$ as follows \cite{PRSA2014}:
  \begin{equation}
\Omega_M(\zeta,\alpha)=(\zeta-\alpha)\prod_{\psi \in \Theta_M''} \frac{(\zeta-\psi(\alpha))(\alpha-\psi(\zeta))}{(\zeta-\psi(\zeta))(\alpha-\psi(\alpha))}.
\label{eq:SKM}
\end{equation}
where $\Theta_M''\subset \Theta_M$ is the set such that for all $\psi\in\Theta_M$, excluding the identity, either  $\psi$ or $\psi^{-1}$ (but not both) is contained in $\Theta_M''$.

The function $\Omega_M(\zeta,\alpha)$ satisfies the following functional identity \cite{PRSA2014}:
\begin{align}
\frac{\Omega_M(\psi_k(\zeta_1),\gamma_1)}{\Omega_M(\psi_k(\zeta_2),\gamma_2)}=
\frac{\tilde \beta_j(\gamma_1,\gamma_2)}{\tilde \beta_j(\zeta_1,\zeta_2)}
\sqrt{\frac{\psi_k'(\zeta_1)}{\psi_k'(\zeta_2)}}
\frac{\Omega_M(\zeta,\gamma_1)}{\Omega_M(\zeta,\gamma_2)}, 
\label{eq:id}
\end{align}
where we recall that $\psi_k=\theta_l\circ\theta_k$, for a fixed $l$ and $k=-M,...,M, k\ne 0,l$, with  $\theta_j(\zeta)$ as defined in (\ref{eq:theta}), are generators of the group $\Theta_M$.  Here
\begin{align}
\tilde \beta_k(\zeta,\gamma)=\exp\left[2\pi i \left(\tilde v_j(\zeta)-\tilde v_j(\gamma)\right)\right],
\end{align}
where the functions $\tilde v_k(\zeta)$  are $2M-1$ integrals of the first kind on the Riemann surface associated with  the Schottky group $\Theta_M$  \cite{Baker}.  
Two particular cases of (\ref{eq:id}) are of interest here. For $\zeta_1=\zeta_2=\zeta$ we have
\begin{align}
\frac{\Omega_M(\psi_k(\zeta),\gamma_1)}{\Omega_M(\psi_k(\zeta),\gamma_2)}=\tilde \beta_k(\gamma_1,\gamma_2) \frac{\Omega_M(\zeta,\gamma_1)}{\Omega_M(\zeta,\gamma_2)}, 
\label{idrat}
\end{align}
while for $\psi_1(\zeta)=\theta_1^2(\zeta)=q_1^4\zeta$ and  $\gamma_1=\gamma_2=\gamma$ we get
\begin{equation}
\frac{\Omega_M( q_1^4\zeta_1,\gamma)}{\Omega_M(q_1^4\zeta_2, \gamma)}=\tilde \beta_1(\zeta_2,\zeta_1)
\frac{\Omega_M( \zeta_1,\gamma)}{\Omega_M(\zeta_2, \gamma)},
\end{equation}
which implies that
\begin{equation}
\frac{\Omega_M( q_1^2\zeta_1,\gamma)}{\Omega_M(q_1^2\zeta_2, \gamma)}=\tilde \beta_1(\zeta_2/q_1^2,\zeta_1/q_1^2)
\frac{\Omega_M( \zeta_1/q_1^2,\gamma)}{\Omega_M(\zeta_2/q_1^2, \gamma)}.
\label{id:t1}
\end{equation}

If we now define the function
\begin{align}
F(\zeta; \gamma_1,\gamma_2) =\log  \frac{\Omega_M(\zeta,\gamma_1)}{\Omega_M(\zeta,\gamma_2)}, 
\label{eq:F}
\end{align}
it then follows  from (\ref{idrat}) that $F(\zeta; \gamma_1,\gamma_2)$ is additive automorphic with respect to the group $\Theta_M$:
\begin{align}
F(\psi_k(\zeta); \gamma_1,\gamma_2) \simeq F(\zeta; \gamma_1,\gamma_2) .
\label{eq:psiF}
\end{align}
where we recall that the equality sign $\simeq$ means that the two functions (on each side of the sign) differ at most by an additive constant. 

The function $\Omega_M(\zeta,\alpha)$  also satisfies the following  symmetry relation \cite{PRSA2014}:
\begin{align}
\overline{\Omega}_M(\phi_j(\zeta),\phi_j(\alpha))=-\frac{q_j^2}{(\zeta-\delta_j)(\alpha-\delta_j)}\, \Omega_M(\zeta,\alpha),
\label{eq:id1}
\end{align}
for $j=0,1,..., M$, where we defined
 \begin{align*}
\overline{\Omega}_M(\zeta,\alpha)\equiv\overline{\Omega_M(\overline\zeta,\overline\alpha)} \, ,
\end{align*}
and $\phi_j(\zeta)$ is the conjugation map on circle $C_j$, i.e., $\bar \zeta=\phi_j(\zeta)$ for $\zeta\in C_j$,  which can be written as
\begin{align}
\phi_j(\zeta)= \overline{\theta}_j(1/\zeta)=\frac{1}{\theta_{-j}(\zeta)}.
\label{eq:phi}
\end{align}
In particular, for $j=0$ relation (\ref{eq:id1}) simplifies to
 \begin{equation}
\overline{\Omega}_M\left(1/\zeta,1/\alpha \right)=-\frac{1}{\zeta \alpha}\Omega_M(\zeta,\alpha).
\label{eq:id2}
\end{equation}
Alternatively, taking the complex conjugate of  (\ref{eq:id1}) and using (\ref{eq:phi}) and (\ref{eq:id2}) we obtain 
\begin{align}
{\Omega}_M(\theta_j( \zeta),\theta_j(\alpha))=\frac{q_j^2}{(1-\overline \delta_j \zeta)(1-\overline\delta_j \alpha)}\, {\Omega}_M(\zeta,\alpha).
\label{eq:id3}
\end{align}
For the particular case $j=1$,  we  have
\begin{align}
{\Omega}_M(q_1^2 \zeta, q_1^2 \alpha)={q_1^2} {\Omega}_M(\zeta,\alpha).
\label{eq:id5}
\end{align}
where we recall that the circle $C_1$  was chosen concentric with the unit circle, so that  $\delta_1=0$ and $\theta_1(\zeta)=q_1^2\zeta$.

\section*{Appendix B: Symmetry Relations for $z(\zeta,t)$}

%As discussed in Sec.~\ref{sec:map}, the map $z(\zeta,t)$  from the circular domain, $D_\zeta$, in the $\zeta$-plane to the fluid region, $D(t)$, exterior to the bubbles in the Hele-Shaw channel satisfies the following boundary conditions on the channel walls:
%\begin{align}
%{\rm Im}\,[z(\zeta,t)]= \mbox{constant}, \quad \mbox{for}\quad \zeta\in C_0.
%\label{eq:Imz}
%\end{align}
%Using that $\overline \zeta=1/\zeta$ for $\zeta \in C_0$,  we can recast   (\ref{eq:Imz}) as
%\begin{align}
%\overline{z}(1/\zeta,t)\simeq z(\zeta,t),
%\label{eq:barz}
%\end{align}
%which is valid for $\zeta\in C_0$ and elsewhere by analytic continuation.

%Let us  consider the boundary conditions on the Schwarz function. 
The Schwarz function ${\cal S}_j(z,t)$ of an interface $\partial D_j$  has the following  representation in the $\zeta$-plane:
\begin{align}
	 g_j(\zeta,t)\equiv {\cal S}_j(z(\zeta,t),t) =\overline {z(\zeta,t)}= \overline z(\overline \zeta,t ) , \quad \zeta\in C_j.
\end{align}
Upon using that $\bar \zeta=\phi_j(\zeta)$ for $\zeta\in C_j$ and  (\ref{eq:phi}),  one obtains
\begin{align}
	g_j(\zeta,t)  = \overline z(1/\theta_{-j}(\zeta),t ) ,
		\label{eq:Sgj}
\end{align} 
which is  valid for $\zeta \in C_j$ and elsewhere by analytic continuation.
Using (\ref{eq:barz})  into (\ref{eq:Sgj}), we then get
\begin{align}
 g_j(\zeta,t) = z(\theta_{-j}(\zeta),t ) ,
\label{eq:SC1}
\end{align}
which recovers (\ref{eq:S}). 

Next, it follows from (\ref{eq:St}), and the boundary conditions on $w(z,t)$, that  ${\cal S}_j(z,t)$ must have constant imaginary part on  the channel walls:  ${\rm Im}[{\cal S}_j(z,t)]=\mbox{constant}$,  for $y=0,\pi$. This implies in turn that ${\rm Im}[g_j(\zeta,t)]=\mbox{constant}$ for $\zeta\in C_0$, or more compactly
\begin{align}
 \overline{g_j(\zeta,t)} \simeq g_j(\zeta,t)  , \qquad \zeta\in C_0.
\label{eq:gjbar}
\end{align} 
Considering the particular case $\zeta\in C_0$ in  (\ref{eq:SC1}) and using (\ref{eq:phi}) and (\ref{eq:barz}), we obtain that  
\begin{align}
 \overline{ g_j(\zeta,t) }  =  \overline z(1/ \theta_{j}(\zeta),t )  =z(\theta_j(\zeta),t ),  \qquad \zeta\in C_0.
\label{eq:SC0}
\end{align}
Combining (\ref{eq:gjbar}) and (\ref{eq:SC0}) we thus have that
\begin{align}
	g _j(\zeta,t) \simeq z(\theta_{j}(\zeta),t ), 		
	\label{eq:SCj}
\end{align}
which  is  valid on $C_0$ and elsewhere by analytic continuation. This proves relation (\ref{eq:Sj}).

Since  different  branches, ${\cal S}_j(z,t)$, of the  Schwarz function  can differ only by an additive constant, see (\ref{eq:glm}),  we can combine (\ref{eq:SC1}) and (\ref{eq:SCj}) as a single general relation:
\begin{align}
	 z(\theta_j(\zeta),t) \simeq z(\theta_{-k}(\zeta),t ) .
	 \label{eq:z4a}
\end{align}
Equivalently, we have
\begin{align}
	 z(\theta_j\circ\theta_k(\zeta),t) \simeq z(\zeta,t ).
		\label{eq:zrho}
\end{align}
Since $\psi_k = \theta_l\circ\theta_k$, for a fixed $l$, are generators of the group $\Theta_M$, see (\ref{eq:psi}), we thus conclude that 
\begin{align}
	 z(\psi(\zeta),t) \simeq z(\zeta,t ), \qquad \psi\in \Theta_M ,
\end{align}
thus showing that $z(\zeta,t )$ is an additive automorphic function with respect to $\Theta_M$, as anticipated in Sec.~\ref{sec:S}.

\section*{Appendix C: Proof of (\ref{eq:pj})}

To show that the complex potential $W(\zeta,t)$  given in (\ref{eq:W}) has constant real part on the circles $C_j$, $j=1,..,M$, let us first consider the case $j=l$. Using that  $\overline \zeta =\phi_l(\zeta)$ for  $\zeta\in C_l$, and the fact $\phi_j(\zeta)= \overline{\theta}_j(1/\zeta)$,   we obtain that 
\begin{align*}
\overline{\left(\frac{\Omega_M(\zeta, \theta_l(\zeta_+))}{ \Omega_M(\zeta, \theta_l(\zeta_-))}\right)}&= 
\frac{\overline \Omega_M(\phi_l(\zeta), \phi_l (\zeta_+))}{ \overline\Omega_M(\phi_l(\zeta), \phi_l (\zeta_-))}\cr
&=\left(\frac{ \delta_l-\zeta_-}{ \delta_l -\zeta_+}\right)\frac{\Omega_M(\zeta, \zeta_+)}{\Omega_M(\zeta, \zeta_-)},
\end{align*}
%for $\zeta\in C_l$, 
where in the last passage we used (\ref{eq:id1}). Inserting this relation into (\ref{eq:W}) we have
\begin{align}
W(\zeta,t) &= \log \left[ \overline{\left(\frac{ \delta_l-\zeta_-}{ \delta_l -\zeta_+}\right) } \frac{Z}{\overline Z} \right]+ ic_l, \qquad \zeta\in C_l,
\label{eq:WApp}
\end{align}
where 
\[ Z= \frac{\Omega(\zeta, \zeta_-)}{\Omega(\zeta,  \zeta_+)}.\] 
It then follows  from (\ref{eq:WApp}) that
\begin{equation*}
p_l(t)=-{\rm Re}\left[ W(\zeta,t)\right]\Big|_{\zeta\in C_l} = \log \left|\frac{ \delta_l -\zeta_+}{ \delta_l-\zeta_-}\right|,
\end{equation*}
thus obtaining (\ref{eq:pj}). 

Similarly, noting that $\theta_l = \psi_{-j} \theta_j$, where $\psi_{-j}=\theta_l\circ \theta_{-j}$ is a generator of the group $\Theta_M$ [see (\ref{eq:psi})], and using the transformation property (\ref{idrat}), one  can show that for $\zeta \in C_j$, $j\ne l$,  one has 
\begin{equation*}
p_j(t)=-{\rm Re}\left[ W(\zeta,t)\right]\Big|_{\zeta\in C_j} = \log \left|\frac{ \delta_j -\zeta_+}{ \delta_j-\zeta_-}\right| +2\pi{\rm Im}[\tilde v_{-j}(\theta_j(\zeta_-))- \tilde v_{-j}(\theta_j(\zeta_+))],
\end{equation*}
where $\tilde v_{-j}(\zeta)$ is the integral of first kind associated with the map $\psi_{-j}$. This result, together with (\ref{eq:pj}),  thus shows that the complex potential $W(\zeta,t)$ has constant real part on all circles $C_j$, $j=1,...,M$, as required.


\begin{thebibliography}{99}
\bibitem{footnote_cl} The expression for $c_l(t)$ is  not given here, as it is irrelevant for the velocity field.
\bibitem{footnote2} We do not present the expression for $\Delta$, as it is irrelevant for the  purposes of this article.
\end{thebibliography}
\end{document}